\documentclass[twocolumn,preprintnumbers,nofootinbib,prl]{revtex4}

\usepackage{graphicx}

\usepackage[hypertex]{hyperref}
\usepackage{amsmath, amssymb}
\newcommand{\vev}[1]{\langle {#1} \rangle}
\newcommand{\lsim}{\lesssim}
\newcommand{\ord}[1]{\mathcal{O}{(#1)}}
\newcommand{\eq}[1]{Eq.~(\ref{#1})}
\newcommand{\re}[1]{Ref.\cite{#1}}
\newcommand{\gsim}{\gtrsim}
\newcommand{\beq}{\begin{equation}}
\newcommand{\eeq}{\end{equation}}

\begin{document}

\pagestyle{plain}


\title{Dark Matter with Time-Varying Leptophilic Couplings}

\author{Hooman Davoudiasl\footnote{email: hooman@bnl.gov} }

\affiliation{Department of Physics, Brookhaven National Laboratory,
Upton, NY 11973, USA}


\begin{abstract}

Two general problems arise when
interpreting the recent cosmic ray data
as signals of Dark Matter (DM) annihilation:
{\it(i)} the required cross section is too large by $\ord{100}$, and
{\it (ii)} the annihilation products seem to be mostly leptonic.  We
propose to address these two problems by assuming that the couplings 
of DM to leptons grow with time.  This can be achieved by
a dynamic localization of DM in extra dimensions.  A possible 
outcome of this proposal is a time (red-shift) dependent
annihilation signal, in terms of strength and dominant final states.

\end{abstract}
\maketitle


The evidence for the presence of Dark Matter (DM), a weakly
interacting cosmic ingredient which is stable on Hubble time scales,
has been gaining strength over the past several decades.  A
confluence of data from rotation curves of galaxies, observations of
Cosmic Microwave Background (CMB), and other large scale surveys of the
universe has established that DM makes up about $25\%$ of the energy
budget of the cosmos \cite{PDG}.  No plausible candidate exists within the
Standard Model (SM) of particle physics for DM, making it one of the
most compelling pieces of evidence that require an extension of our
current microscopic understanding of Nature.
Among the possible candidates for DM, Weakly Interacting
Massive Particles (WIMP's) are generic in models
that address the stability of the Higgs
potential against large quantum correction, or the ``hierarchy"
problem.  Here, we will focus on this class of particles.  It turns
out that the weak scale of $\ord{100}$~GeV sets the correct level of
DM annihilation cross section in the early universe, resulting in an acceptable
thermal relic abundance.

The search for DM is carried out in several ways, encompassing
direct laboratory detection and indirect methods which rely on the
annihilation signals of DM.  Of these latter signals, possible hints
in cosmic ray data have recently received a great deal of attention.
In particular, the electron flux anomaly observed by the ATIC experiment \cite{:2008zzr}
and the excess positron fraction measured by the PAMELA
satellite \cite{Adriani:2008zr} have been subject of a flurry of activity.  There are
other experiments that have presented signals, potentially consistent with
observations of DM.  However, we will only concentrate on the aforementioned
cosmic ray data in the $\ord{100}$~GeV range and will not attempt to account for any other
signals. In particular, we will focus on two apparent problems that
arise in interpreting the measurements as pointing to WIMP
annihilation (we note that 
purely astrophysical $e^+ e^-$ production 
by nearby pulsars has also been considered as an 
explanation of these data in, 
for example, Refs.~\cite{Hooper:2008kg,Profumo:2008ms}; 
though this remains an interesting possibility, we will 
discuss only a WIMP-based solution here): 
{\it (i)} the required annihilation cross section is
too large by $\ord{100}$ \cite{Cholis:2008hb}, 
if one assumes that DM density was set in
the early universe, when similar annihilation processes froze out,
and {\it (ii)} assuming that DM is a weak scale particle, it is
puzzling that the cosmic ray data do not reveal an excess in the
anti-proton flux.

Recently, {\it (i)} and {\it (ii)} have been addressed, for example, by postulating a
light degree of freedom that couples to DM and mediates a long range
force \cite{ArkaniHamed:2008qn}, or in the context of 
the Stueckelberg mechanism \cite{Feldman:2008xs}.
These are interesting possibilities that deserve
attention, and they may also lead to distinct signals at the Large
Hadron Collider \cite{ArkaniHamed:2008qp,Feldman:2008xs}.
However, we note that, in these proposals,
the mass scales of the WIMP and the mediator have to be chosen
judiciously such that the desired effect is obtained.  The
resolution to question {\it (ii)} above has also been discussed
in models of leptophilic WIMP \cite{leptophil}.

In this work, we propose that both {\it (i)} and {\it (ii)} can be
addressed by assuming a time-varying leptophilic coupling of the
WIMP.  In our proposal, we do not constrain the early universe
couplings of DM to be strictly leptophilic and only require this to
be a late-time feature.  In fact, in our framework, the primordial
couplings of DM to the SM can even be baryophilic and leptophobic.
The main requirement is that DM couples predominantly to $e^\pm$ or $\mu^\pm$ 
today and that these or other SM couplings were somewhat smaller at freeze out.  
This is the basic idea that we present here.
However, we will discuss a specific realization of this scenario in
the context of models with extra dimensions
\cite{EXD}.  One may place the SM
field content in some of the extra dimensions.
In particular, fermions can be localized
at different points along the compact dimensions, giving rise to
suppressed coupling among various fermions, based on their location
\cite{splitfermions}.  Here, we note that DM with time varying
properties has been considered in Ref.~\cite{Anderson:1997un}, and
more recently in Ref.~\cite{Cohen:2008nb}.  Dynamic
fermion localization, in the context of baryogenensis, has been
studied in Refs.~\cite{Masiero:2000fr,Chung:2001tp,Perez:2005yx}.

On general grounds, any time variation in the DM couplings can lead to
red-shift dependent annihilation signals.  In our proposal, one
would expect that lepton final states become more dominant at
smaller red-shifts.  Of course, this depends on the exact rate of
time variation in the leptophilic couplings.  Depending on the specific
details, one could end up with the early halo annihilations that
are dominated by hadronic final states; here, galaxies at high
red-shift may give distinctly different signals.  In certain cases,
we could also have a situation in which dark matter
annihilation becomes inefficient over a non-negligible cosmological
epoch after the early freeze-out.  Finally, in the
extra dimensional scenario that we consider, the
hierarchy problem motivates compactification radii of $\ord{1/{\rm
TeV}}$ or larger, with the usual attendant collider signatures.

We will adopt the model presented in \re{splitfermions}, as a
concrete example, and for simplicity,
will focus on the fermion content in one extra
dimension $y$ of size $R \gsim$~1/TeV.
The field content considered here will
suffice to demonstrate our main ideas.
We will make no attempt at making the setup fully realistic, lest
the key features get lost in unnecessary details.  Further components
can be added to our model in a more expansive work.

Let $\chi$ denote a singlet fermion
and $S$ a singlet scalar.
Here, $\chi$  gets localized along the extra
dimension by its coupling with the bulk profile of $S$, via
the 5D Lagrangian
\beq
{\cal L}_5 = \int d^4 x\, d y\,
{\bar \chi} (i\partial\!\!\!\slash + S_\chi(y) - m_{5\chi}) \chi,
\label{L5}
\eeq
where $\partial\!\!\!\slash \equiv \gamma^\mu \partial_\mu + \gamma^5 \partial_5$ 
is the 5D derivative and $m_{5\chi}$ is a bulk Dirac mass for $\chi$.
Following \re{splitfermions}, we assume the simple harmonic oscillator
profile $S_\chi(y) = 2\mu_\chi^2 y$.   However, this profile ultimately
reaches constant values, giving rise to a wall of thickness $\Delta\lsim R$ \cite{splitfermions}.
The resulting fermion wave-function $f_\chi$ for the zero mode of $\chi$ is a
Gaussian centered at $y_\chi = m_{5\chi}/(2 \mu_\chi^2)$
\beq
f_\chi(y) = \left(\frac{\mu_\chi}{\sqrt{\pi/2}}\right)^{1/2} e^{-\mu_\chi^2(y - y_\chi)^2}.
\label{fchi}
\eeq

Let $\phi$ be a 5D scalar and a doublet of the SM $SU(2)$ that
does not develop a vacuum expectation value (vev); $\vev{\phi}=0$.
We will assume that both $\chi$ and $\phi$ are
charged under a ${\mathbb Z}_2$ symmetry.  The
following 5D operator yields the leading interaction between $\chi$ and $\phi$
\beq
O_1 = \lambda_\phi \sqrt{\Delta}\, \chi\, \phi L ,
\label{O1}
\eeq
where $\lambda_\phi \sqrt{\Delta}$ is a 5D Yukawa coupling \cite{Mirabelli:1999ks};
$L$ is a lepton doublet of the SM.  One can easily localize the fermions
of the SM, using straightforward extensions of \eq{L5}.  We note that in order
to identify $\chi$ as a DM candidate, we will assume that the {\it zero mode
masses} satisfy $m_\phi > m_\chi$.  We may include 
a term $m_\phi^2 \phi^\dagger \phi$ in the 
5D action to endow $\phi$ with mass $m_\phi \gsim 10^2$~GeV.  
The mass of the zero mode for $\chi$ may
be induced in a variety of ways.  For now, we assume that a Majorana mass
$m_\chi \chi^T C \chi$, with $C$ the 5D charge conjugation operator,
can be induced in the bulk.  Given the typical expectation
$m_\chi \sim 100$~GeV, this is a small mass compared to other scales
and we treat it as a perturbation.

We will further assume that the profile of the SM Higgs doublet 
is flat over the width of the wall.
Bulk masses $m_{5f}$ of the SM fermions allow one to localize them at various
points $y_f$ along the fifth dimension.  Given the exponential nature of the
fermion wave-functions, a wide range of 4D masses $m_f$ can be obtained without
recourse to small Yukawa couplings;
$
m_f \sim e^{-(\mu_f s)^2/2}\, v_{EW}
$ \cite{splitfermions}.
Here, $v_{EW}\sim 10^2$~GeV is the Higgs vev and $s$ is the separation of the SM $SU(2)$
doublet and singlet fields along the fifth dimension; we have assumed $\ord{1}$
Yukawa couplings and a separate
localizing profile for the SM fields, characterized by $\mu_f$, which is not strictly
necessary.  In this way, the lightest
fermions are typically the farthest separated.
We will mainly focus on lepton localization.

The proximity of $\chi$ to the first generation lepton doublet
$L_e$, at the present epoch, ensures a significant coupling only to
these fermions. If the the heavier charged lepton doublets
$L_{\mu,\tau}$ are mildly separated from $L_e$, they would not
couple to the DM candidate $\chi$ significantly.  This feature can
naturally account for $\chi \chi \to e^+ e^-$ as a preferred
$\phi$-mediated present-day-annihilation channel.  Note that in this
particular case, the dominant coupling is to the left handed fields
$(e, \nu_e)_L$, and hence annihilation into $\nu_e$ pairs (or other 
neutrino flavors, corresponding to different dominant 
charged lepton final states) can be comparable to that in the $e^+ e^-$ channel.  
This could lead to possible experimental signatures in astrophysical high energy neutrino 
experiments; see, for example, Ref.~\cite{Spolyar:2009kx} for a discussion.

With a roughly flat zero mode profile for $\phi$,
the annihilation of $\chi$ is dominantly governed by the coupling
\beq
g_{\chi e} \sim \sqrt{\frac{2 \mu_e \mu_\chi}{\mu_e^2 + \mu_\chi^2}}\,
e^{- \omega^2 s_{\chi e}^2},
\label{gchie}
\eeq
where
$
\omega^2\equiv (\mu_e^2 \mu_\chi^2)/(\mu_e^2 + \mu_\chi^2)
$
and $s_{\chi e} = |y_\chi-y_e|$ is the separation between the Gaussian centers;
different localizing scalars for $L_e$ and $\chi$ are assumed.

If we keep the masses of the particles fixed,  the annihilation cross section of $\chi$
is of order $\vev{\sigma_a v} \sim g_{\chi e}^4/m_\phi^2$, with $m_\chi \lsim m_\phi$.
Hence, getting a factor of $\ord{100}$ enhancement in
the cross section at late times roughly requires $g_{\chi e} \to 3 g_{\chi e}$, approximately
one $e$-folding, which only amounts to an $\ord{1}$
shift of the exponent in \eq{gchie}.  Before discussing the required shift, let us
simplify \eq{gchie} by assuming that all fermions are localized by the same
profile and $\mu_\chi = \mu_e=\mu$.  We then have,
\beq
g_{\chi e}\sim e^{-(\mu \,s_{\chi e})^2/2}.
\label{gchiesimp}
\eeq
Thus, we need $s_{\chi e}^2 \to s_{\chi e}^2 -2/\mu^2$.  Since we are only interested
in the relative distance between the fermions, let us assume, for purposes of demonstration,
that $y_e = 0$.  Then, we find that the new position of $\chi$ should be at $y^\prime_\chi =
m^\prime_{5\chi}/(2\mu^2)$, where
\beq
m^\prime_{5\chi} = \sqrt{m_{5\chi}^2 - 8\mu^2}.
\label{m5chiprime}
\eeq
Note that accommodating a realistic set of masses in the SM typically
requires $\mu s$ to be of order a few \cite{splitfermions}.  Hence, one could typically expect
$m_{5f}/\mu \gg 1$, leading to
$m^\prime_{5\chi} \approx m_{5\chi}(1 - 4\mu^2/m_{5\chi}^2)$.  Thus, we estimate
\beq
|\delta m_{5\chi}| \lsim \mu.
\label{delm5}
\eeq
This relation quantifies the shift in the value of $m_{5 \chi}$, in
terms of the localizing potential parameter $\mu$,
required to account for both the primordial and present day annihilation
cross section of $\chi$.  We have implicitly assumed that both rates
are dominated by the interaction in \eq{O1}.  This in turn implies that other
lepton doublets did not have a major overlap with $\chi$,
during the evolution of $g_{\chi e}$.

Let us now estimate what \eq{delm5} entails in some motivated scenarios.  Given that
$\Delta$ would set the typical mass scale of Kaluza-Klein (KK) modes of SM particle trapped
within the wall, we expect $\Delta^{-1} \gsim 10^2$~GeV \cite{splitfermions}; we will take
$\Delta^{-1} \sim 1$~TeV.  Following Ref.~\cite{splitfermions}, we may then expect
$\mu \sim 10 \Delta^{-1}\sim 10$~TeV.  So far, we have assumed the 5D mass parameter for
fermions, and in particular $\chi$, to be a Lagrangian parameter.  However, in general, we
may postulate that this mass is set by the vev of a 5D scalar $\Phi$ (note that this field 
is not that in Ref.~\cite{splitfermions} responsible for localizing the fermions).  Assuming a constant
rate of change of $\vev{\Phi}$ over the cosmological time scale $t_0\sim H_0^{-1}$, where
$H_0 \sim 10^{-33}$~eV is the present day Hubble constant, and $\delta \Phi \sim \mu$
we find
\beq
\dot{\Phi}^2 \sim (\mu H_0)^2 \sim 10^{-40}~{\rm eV}^4.
\label{Phidot2}
\eeq

The inferred cosmic dark energy density is $\ord{10^{-11}}$~eV$^4$ \cite{PDG}.
So, we do not expect this rate of change for $\Phi$ to have a significant effect on the
cosmological evolution of the universe.  Nonetheless, obtaining such a tiny rate of change for
$\Phi$ is fraught with difficulties, as this field must have an extremely small mass,
$m_\Phi \lsim H_0$, so as to vary over cosmic time scales (a familiar feature in
quintessence models).  As we need $\Phi$-$\chi$ coupling of $\ord{1}$, maintaining such a
fine-tuned potential poses a theoretical challenge.  Also, a nearly massless
scalar with significant couplings to DM could cause violations of equivalence principle,
which is potentially another undesirable feature.  Here, we instead relegate the variation
of $m_{5\chi}$ to a temporal change in the coupling $\xi$ of $\Phi$ to $\chi$
and hold $\vev{\Phi}$ fixed.  We will consider cases where $\xi$ depends on 
dynamical moduli in the gravitational sector.  The couplings of such moduli are typically 
suppressed by the Planck mass $M_P\sim 10^{19}$~GeV, which in turn greatly suppresses   
loop-induced contributions to their masses \cite{Chacko:2002sb}, 
as well as possible undesirable light-field-mediated effects.

Let $m_{5\chi} = \xi \vev{\Phi}$.  As that the field $\Phi$ is a gauge singlet,
we may allow it to travel in various dimensions.  It is then reasonable to assume that
its coupling to $\chi$ depends on the volume of extra dimensions.  We will adopt the ansatz
\beq
\xi \propto V_n^{-1/2},
\label{xi}
\eeq
with $V_n$ the volume of the $n$ extra dimensions where $\Phi$ can propagate.  From our previous
discussion, we may expect that $\delta m_{5\chi} \sim - m_{5\chi}/10$, which yields
$\delta V_n/V_n \sim 0.1$, ignoring factors of $\ord{n}$.  Hence, we require that the volume of
some extra dimensions change by about $10\%$ over a Hubble time.

We will not speculate on the nature of the potential that gives
rise to the aforementioned slow evolution.  However, we point out that the initial
change of $\xi$ cannot be too fast, as it could yield unwanted consequences.  First of all,
if this change takes place too quickly over the time scales of freeze out process, one may end up with
insufficient amounts of DM.  To avoid this problem, we then require that at time
$t_{fo} \sim M_P/T_{fo}^2$, where $T_{fo}$ is the freeze out temperature, $\delta \xi/\xi \ll 0.1$.  We also
require that the predictions of Big Bang Nucleosynthesis (BBN) be largely unperturbed.  This implies that
the kinetic energy in the radion $(\partial_t r)^2$
must not be the dominant energy density, so that a radiation
dominated cosmological scenario persists
(see Ref.~\cite{DeFelice:2005bx} for an alternative
setup).  Assuming a typical BBN temperature of
$T\sim$~MeV, during this era ($t\sim 1$~s), we then roughly require
\beq
\partial_t r\lsim {\rm MeV}^2\quad {\rm (BBN)}.
\label{BBN}
\eeq

Here, we note that there is a map between the radion, corresponding to the
volume of the extra dimensions, and a Brans-Dicke (BD) field; see for example Ref.~\cite{Cline:2002mw}.
The BD parameter $\omega \to 1/N - 1$, where $N$ is the total number of extra dimensions and we have
$n \leq N$.  The agreement of various observations with standard cosmology, governed by general relativity,
then places constraints on the value of $\omega$.
Solar system measurements require $\omega > 40000$ today \cite{Bertotti:2003rm}.
However, there are also stringent bounds that come from
WMAP observations of the CMB
and large scale structure, suggesting that $|\omega| \gsim 100$ \cite{Wu:2009zb}.
Clearly, this is in conflict with having a dynamical radion field that varies
over a significant fraction of Hubble time.  To go further, we then consider
a few possible scenarios.

Since we are only interested in shifting the couplings of $\chi$ to
the SM after freeze out, it is, strictly speaking, unnecessary to
assume any time variation long after this era.  In generic WIMP
scenarios, $T_{fo} \sim 10$~GeV, corresponding to $t_{fo} \sim
10^{-5}$~s.  This is well before the CMB decoupling era ($t\sim 10^5$~yr), and even
that of BBN.  However, only signals related to extra dimensions,
and not cosmological signatures of time variation, would likely be
observable in this case.  Even though this is a consistent option,
we will not pursue it further here.

Another possibility is suggested upon noting that the BD constraints
obtained in \re{Wu:2009zb} are in fact mostly a result of
constraints on the time variation of Newton's constant $G_N$.  This
variation, in our setup, is a consequence of change in the volume of
the extra dimensions.  Note that we are not necessarily assuming
that $\xi$ depends on the total volume of the extra dimensions,
which is governed by the radion potential; we only assume that the
coupling $\xi$ depends on {\it some} of the radii. If some radii
change in a way that ensures the volume of the extra dimensions
remains constant over cosmological times \cite{Cline:2002mw} we may
bypass the constraints on variation of $G_N$ and hence the BD
parameter $\omega$.  This will correspond to a Kasner-like behavior
for the cosmology, where some directions expand while others
contract.  We will assume that such a solution can be constructed,
leading to a constant extra-dimensional volume, while allowing some
of the radii to vary with time.  As long as $\xi$ depends on a
subset of such radii, we will achieve a situation where $\xi$, and
therefore the localization of $\chi$, can change over times
of $\ord{H_0^{-1}}$, without an appreciable change in $G_N$.
This could lead to models that are unencumbered by the bounds on
$\omega$, as assumed in the following.

The basic requirements of our scenario leave a wide
range of possibilities open to model building.  As an alternative
example, we will outline a setup in
which $\chi$ starts out baryophilic, having $\ord{1}$
couplings to the quark sector. Over cosmological times, the
evolution of $\xi$ changes the localization of $\chi$,
moving it towards leptons.  At late times, $\chi$ ends
up being leptophilic and baryophobic.

The above possibility, in fact, fits very well with one of the
original motivations for fermion localization in extra dimensional
models, that is the geometric suppression of proton decay
\cite{splitfermions}, by separating leptons and quarks by a
sufficient distance over the width of the wall.
Even if the cutoff of the theory is $M_F\gsim 1$~TeV,
one can show that a separation $s \sim 10/\mu$ will suffice to
get a safely long-lived proton \cite{splitfermions}.
If $\chi$ mainly couples to quarks initially, we need a new state with the right
quantum numbers to provide the necessary annihilation cross section.
For example, a new scalar $\theta$ with the conjugate quantum
numbers of $SU(2)$ singlet $u_R$ quarks will in principle be
sufficient.  As for $\phi$ in the previous case, we need to assume
that $\theta$ is charged under the same ${\mathbb Z}_2$ symmetry as
$\chi$, $\vev{\theta}=0$, and $m_\theta > m_\chi$.  In analogy with \eq{O1},
$\theta$ couples via
\beq
O_2 = \lambda_\theta \sqrt{\Delta}\, \chi\, \theta u_R ,
\label{O2}
\eeq
where $\lambda_\theta \sqrt{\Delta}$ is a 5D Yukawa coupling.

Given a separation $s \sim 10/\mu$
of leptons and quarks required to protect the proton from
fast decays, we now require $\delta m_{5 \chi} \sim 20 \mu$, after
freeze out.  Hence, some of the estimates from
the above discussion will be modified here, but not enough
to change the qualitative conclusions of the previous case.  As the field
$\chi$ moves toward its current location, one can expect its
couplings to various SM fields to rise and fall with time, in a
model dependent manner.  However,
with separated quarks and leptons in well-motivated models,
a few general features may be expected to arise.  

If the evolution of
couplings takes place over cosmological times,
we may start with a population of DM that mostly annihilates into quarks.  This
will entail a hard photon signal component 
coming from $\pi^0 \to \gamma \gamma$ at high
redshifts.  To get a rough estimate of the possible signal, let us assume that 
the photons originate from a galaxy cluster of size $l_c \sim 1$~Mpc at a distance 
$D_c$ from the earth.  We take a DM number density $n_\chi \sim 10^{-3}$~cm$^{-3}$ 
and an annihilation cross section $\vev{\sigma_a v} \sim 10^{-23}$~cm$^3$/s, similar 
in size to the implied leptonic cross section at the present epoch.  The annihilation rate 
into $q {\bar q}$ is then given by 
\beq
\Gamma_a \sim l_c^3 \, n_\chi^2  \vev{\sigma_a v} \sim 10^{44} \,{\rm s}^{-1}.
\label{Gam_a}
\eeq  
Given that at $\ord{100}$~GeV energies we can expect an $\ord{10}$ multiplicity 
of neutral pions \cite{PDG}, the flux 
$f_\gamma$ of photons at earth is approximately 
given by 
\beq
f_\gamma \sim \Gamma_a/D_c^2 \sim 10^{-9}\left(\frac{300~{\rm Mpc}}{D_c}\right)^2\, 
{\rm cm}^{-2} {\rm s}^{-1}.
\label{f_gam}
\eeq
Hence, if the quark annihilation channel was dominant around $10^9$ years ago, 
Fermi gamma-ray space telescope may be able to detect \cite{Fermi-SRD} the photons from
$\pi^0 \to \gamma \gamma$ at the corresponding redshift. 

As $\chi$ enters a `fermionic desert' between the loci of quarks
and leptons, its major couplings are governed by higher dimensional operators
that are suppressed by powers of the cutoff scale $M_F \gg \Delta^{-1}$.  Hence,
one can expect a drop in the efficiency of DM annihilation over intermediate
red-shifts.  Once $\chi$ moves closer to the locus of leptons along the extra dimension,
its couplings become leptophilic and larger, leading to enhanced DM annihilation, dominated
by lepton final states, at lower redshifts.

In this paper, we proposed that the observed
cosmic ray excess in positrons, and not anti-protons, can be due to
the annihilation of DM with time-varying couplings that
become more leptophilic after freeze out.  A
possible consequence of this scenario is a red-shift dependent
annihilation signal.  We discussed how these features may be
incorporated in a model where the SM fermions and the DM fermion
$\chi$ are localized along an extra dimension. If the localization
of $\chi$ moderately changes after freeze out, it could lead to
enhanced coupling to electrons and hence explain the recent data.
Depending on details, the annihilation at earlier epochs and high
redshifts could in fact be dominated by quark final states, resulting in hard
photons from $\pi^0$ decays.  

{\it Note added:} After this work was posted, a new data set  
from the Fermi collaboration on the $e^+ + e^-$ cosmic ray spectrum 
\cite{Abdo:2009zk} was released.  The new measurements do not 
confirm the excess flux around 500~GeV reported by 
the ATIC collaboration \cite{:2008zzr}, but they still allow 
a DM interpretation.  In particular, a DM candidate with a mass of a few 100~GeV, 
leptophilic couplings, and an enhanced annihilation 
cross section at the present epoch remains a possibility \cite{Grasso:2009ma} 
for which our work proposes a time-varying extra dimensional scenario.

\acknowledgments

We thank Mark Trodden for discussions. This work was supported by
the United States Department of Energy under Grant Contract
DE-AC02-98CH10886.

\end{document}